\documentclass[prd,twocolumn,tightten,aps,nofootinbib,showpacs,preprintnumbers]{revtex4}
\begin{document}
\title{On the interplay between screening and confinement from interacting
electromagnetic and torsion fields}
\author{Patricio Gaete}
\email{patricio.gaete@usm.cl} \affiliation{Departmento de
F\'{\i}sica and Centro Cient\'{\i}fico-Tecnol\'{o}gico de
Valpara\'{\i}so,Universidad T\'ecnica Federico Santa Mar\'{\i}a,
Valpara\'{\i}so, Chile}
\author{Jos\'{e} A. Hela\"{y}el-Neto}
\email{helayel@cbpf.br} \affiliation{Centro Brasileiro de Pesquisas
F\'{\i}sicas, Rua Xavier Sigaud, 150, Urca, 22290-180, Rio de
Janeiro, Brazil}
\date{\today}

\begin{abstract}
Features of screening and confinement are studied for the
coupling of axial torsion fields with photons in the presence of an
external electromagnetic field. To this end we compute the static
quantum potential. Our discussion is carried out using the
gauge-invariant but path-dependent variables formalism which
is alternative to the Wilson loop approach. Our results show that,
in the case of a constant electric field strength expectation value,
the static potential remains Coulombic, while in the case of a
constant magnetic field strength expectation value the potential
energy is the sum of a Yukawa and a linear potential, leading to
the confinement of static probe charges.
\end{abstract}
\maketitle

\section{Introduction}

The formulation and possible experimental consequences of extensions of
the Standard Model (SM) such as torsion fields have been vastly investigated
over the latest years  \cite{ShapiroNPB,ShapiroPR,Camelia,Liberati,Hehl,Audretsch,Rumpf,Helayel,Moon}.
As is well-known, this is because the SM does not include a quantum
theory of gravitation. In fact, the necessity of a new scenario has
been suggested to overcome difficulties theoretical in the quantum
gravity research. In this respect we recall that string theories
\cite{Witten} provide a consistent framework to unify all fundamental
interactions. We also point out that string theories are endowed with
interesting features such as a metric, a scalar field (dilaton) and an
antisymmetric tensor field of the third rank which is associated with
torsion. It is worth recalling at this stage that, in addition to the
string interest, torsion fields have been discussed under a number of
different aspects. For instance, in connection to the observed
anisotropy of the cosmological electromagnetic propagation
\cite{Nodland,Maroto}, the dark energy problem \cite{Minkevich},
and in higher dimensional theories \cite{German,Oh}. The ongoing
activity of the Large Hadronic Collider (LHC) has also attracted
interest in order to test the dynamical torsion parameters
\cite{Belayev} and, related to this issue, the production of light
gravitons \cite{Hagiwara,Ya,Cartier,Jain} at accelerators justifies
the study of dynamical aspects of torsion.

On the other hand, in recent times the coupling of axial torsion fields
with photons in the presence of an external background electromagnetic
field and its physical consequences have been discussed \cite{Kruglov}.
Meanwhile, in a previous paper \cite{GaeteHel09}, the impact of
axial torsion fields on physical observables  in terms of the
gauge-invariant but path-dependent variables formalism has been studied.
Specifically, we have computed the static potential between test charges
for a system consisting of a gauge field interacting with propagating
axial torsion fields when there are nontrivial constant expectation values
for the gauge field strength $F_{\mu\nu}$. According to this formalism,
in the case of a constant electric field strength expectation value the
static potential remains Coulombic. Interestingly enough, in the case of
a constant magnetic field strength expectation value the potential energy
is linear, that is, the confinement between static charges is obtained.
In this case, the mass of torsion fields (spin-$1$ states) contribute
to the string tension. The picture which emerges from this study is
that the coupling of torsion fields with photons in the presence of
a constant magnetic field strength expectation, behaves like small
magnetic dipoles in an external magnetic field.

With these considerations in mind, the present work is a sequel to Ref.
\cite{GaeteHel09}. To do this, we will work out the static potential for
a theory which includes both spin-$1$ and spin-$0$ states for the axial
torsion field $S_\mu$ coupled to photons in the presence of an external
background electromagnetic field, using the gauge-invariant but path-dependent
variables formalism. Following this procedure we obtain that, in the case
of a constant electric field strength expectation value the static
potential remains Coulombic. While in the case of a constant magnetic
field strength expectation value we obtain that the potential energy is
the sum of a Yukawa and a linear potential, leading to the confinement
of static charges. This clearly shows the role played by the spin-$0$
state of  the torsion field $S_\mu$ in yielding the Yukawa potential.
It is to be noted that the above static potential profile is analogous
to that encountered in axionic electrodynamics \cite{GaeteG}. Therefore,
the above result reveals a new equivalence between effective Abelian
theories. As well as, the gauge-invariant but path-dependent variables
formalism offers an alternative view in which some features of effective
Abelian theories become more transparent.

\section{Interaction energy}

We shall now discuss the interaction energy between static point-like
sources for the model under consideration. To this end, as in
\cite{GaeteHel09}, we will compute the expectation value of the energy
operator $H$ in the physical state $|\Phi\rangle$ describing the sources,
which we will denote by ${\langle H\rangle}_\Phi$. To carry out our study
we consider the Lagrangian density \cite{ShapiroNPB,Kruglov}:
\begin{eqnarray}
{\cal L} &=&  - \frac{1}{4}F_{\mu \nu }^2  - \frac{1}{4}S_{\mu \nu }^2
+ \frac{1}{2}m^2 S_\mu ^2  - \frac{b}{2}(\partial_\mu S^\mu)^2 \nonumber\\
&+& \frac{g}{4}S^\lambda  \partial _\lambda
\left( {F_{\mu \nu } \tilde F^{\mu \nu } } \right), \label{Tormag5}
\end{eqnarray}
where $S_{\mu \nu }  = \partial _\mu S_\nu   - \partial _\nu  S_\mu$,
${\widetilde F}_{\mu \nu }  \equiv {\raise0.7ex\hbox{$1$}
\!\mathord{\left/{\vphantom {1 2}}\right.\kern-\nulldelimiterspace}
\!\lower0.7ex\hbox{$2$}}\varepsilon _{\mu \nu \lambda \rho }
F^{\lambda \rho }$, $g$ is a coupling constant with dimension
$(-2)$ in mass units, and $b = \frac{{m^2 }}{{m_0^2 }}$. $m$ and
$m_0$, respectively, denote the masses of spin-1 and spin-0 states for
the torsion field ($S_\mu$).

According to the results of the paper of Ref. \cite{Helayel} on the
constraints to be obeyed by quantum torsion, both the spin-$1$ and the
spin-$0$ excitations of $S_\mu$ must be much more massive than the
fundamental particles of the Standard Model. Also, if we assume that the
spin-$0$ mode is much heavier than the spin-$1$ component of $S_\mu$,
the presence of the $(\partial_\mu S^\mu)^2$-term in (\ref{Tormag5})
becomes harmless, in that the ghost mode that would run into troubles
with unitarity is suppressed \cite{Kruglov}. Then, considering the
situation for which
\begin{equation}
m_0^2  \gg m^2  \gg m_{SM}^2, \label{Tormag5a}
\end{equation}
where $m_{SM}^2$ stands for the masses of the fundamental particles of
the Standard Model, we are allowed to integrate over the $S_\mu$-field 
and consider an effective model that describes physics at scales
$\ll m^2$.

Next, by integrating out the $S_\mu$ field in expression
(\ref{Tormag5}), one gets an effective Lagrangian density:
\begin{equation}
{\cal L} =  - \frac{1}{4}F_{\mu \nu }^2  + \frac{{g^2 }}{8}
\left( {F_{\mu \nu } \tilde F^{\mu \nu } } \right)\frac{\Delta }
{{(b\Delta  + m^2) }}\left( {F_{\mu \nu } \tilde F^{\mu \nu } } \right),
\label{Tormag10}
\end{equation}
where $\triangle\equiv\partial^\mu\partial_\mu$. Now, after
splitting $F_{\mu\nu}$ in the sum of a classical background,
$\left\langle {F_{\mu \nu } } \right\rangle$, and a small fluctuation,
$f_{\mu \nu } =\partial _\mu a_\nu -\partial _\nu a_\mu$, the corresponding
Lagrangian density is given by
\begin{equation}
{\cal L} =  - \frac{1}{4}f_{\mu \nu }^2  + \frac{{g^2 }}{8}
\left( {v^{\mu \nu } f_{\mu \nu } } \right)\frac{\Delta }
{{(b\Delta  + m^2) }}\left( {v^{\lambda \gamma } f_{\lambda \gamma } } \right). \label{Tormag15}
\end{equation}
Here, we have simplified our notation by setting
$\varepsilon ^{\mu \nu \alpha \beta } \left\langle{F_{\mu \nu } } \right\rangle
\equiv v^{\alpha \beta }$. This effective theory thus provides us with a
suitable starting point to study the interaction energy. It is
straightforward to check that in the limit $b=0$, expression
(\ref{Tormag15}) reduces to our previous effective Lagrangian \cite{GaeteHel09}.

\subsection{Magnetic case}

We now proceed to obtain the interaction energy in the $v^{0i} \ne
0$ and $v^{ij}=0$ case (referred to as the magnetic one in what
follows). In such a case, the Lagrangian density (\ref{Tormag15})
reads as below:
\begin{equation}
{\cal L} =  - \frac{1}{4}f_{\mu \nu }^2  + \frac{{g^2 }}{8}
\left( {v^{0i} f_{0i} } \right)\frac{\Delta }{{(b\Delta  + m^2) }}
\left( {v^{0k} f_{0k} } \right),
\label{Tormag20}
\end{equation}
where $\mu ,\nu  = 0,1,2,3$ and $i, k = 1,2,3$. We now restrict our attention
to the Hamiltonian framework of this theory. The canonical momenta read
$\Pi ^\mu   = f^{\mu 0}  + \frac{{g^2 }}{4}v^{0\mu }
\frac{\Delta }{{\left( {b\Delta  + m^2 } \right)}}v^{0k} f_{0k}$. This yields
the usual primary constraint  $\Pi^{0}=0$, while the momenta are $\Pi _i  =
D_{ij} E_j$. Here $E_i  \equiv f_{i0}$ and $D_{ij}  = \delta _{ij}  +
\frac{{g^2 }}{4}v_{i0} \frac{\Delta }{{\left( {b\Delta  + m^2 } \right)}}v_{j0}$.
Since ${\bf D}$ is nonsingular, there exists its inverse, ${\bf D}^{-1}$.
With this, the corresponding electric field can be written as
\begin{equation}
E_i  = \frac{1}{{\det D}}\left\{ {\delta _{ij} \det D - \frac{{g^2 }}
{4}v_{i0} \frac{\Delta }{{\left( {b\Delta  + m^2 } \right)}}v_{j0} }
\right\}\Pi _j.  \label{Tormag25}
\end{equation}
Therefore, the canonical Hamiltonian takes the form
\begin{eqnarray}
H_C  &=& \int {d^3 x} \left\{ { - a_0 \partial _i \Pi ^i  +
\frac{1}{2}{\bf B}^2 } \right\} \nonumber\\
&-& \int {d^3 x} \frac{1}{2}\Pi _i \left[ {1 - \frac{{g^2 {\bf v}^2 }}{4}
\frac{\Delta }{{\left( {\xi ^2 \Delta  + m^2 } \right)}}}
\right]\Pi ^i, \label{Tormag30}
\end{eqnarray}
with $\xi ^2  \equiv b + \frac{{{\bf v}^2 g^2 }}{4} = b + {\bf v}^2 g^2
{\cal B}^2$. Here, $\bf B$ and ${\bf {\cal B}}$ stand, respectively, for the
fluctuating magnetic field and the classical background magnetic
field around which the $a^\mu$-field fluctuates. $\bf B$ is
associated to the quantum  $a^\mu$-field: $B^i  =
-\frac{1}{2}\varepsilon_{ijk}f^{jk}$, whereas ${\cal B}_i$,
according to our definition for the background  $\left\langle
{F_{\mu \nu } } \right\rangle$ in terms of $v_{\mu \nu }$ is given
by ${\cal B}_i  = \frac{1}{2}v_{0i}$. Temporal conservation of the
primary constraint $\Pi_0$ leads to the secondary constraint
$\Gamma_1 \left(x \right) \equiv \partial _i \Pi ^i=0$. It is
straightforward to check that there are no further constraints in the
theory. Consequently, the extended Hamiltonian that generates translations
in time then reads $H = H_C + \int {d^3 }x\left( {c_0 \left( x
\right)\Pi _0 \left( x \right) + c_1 \left( x\right)\Gamma _1 \left(
x \right)} \right)$. Here $c_0 \left( x\right)$ and $c_1 \left( x
\right)$ are arbitrary Lagrange multipliers. Moreover, it follows from
this Hamiltonian that $\dot{a}_0 \left( x \right)= \left[ {a_0
\left( x \right),H} \right] = c_0 \left( x \right)$, which is an arbitrary
function. Since $\Pi^0 = 0$ always, neither $ a^0 $ nor $ \Pi^0 $ are of
interest in describing the system and may be discarded from the theory.
As a result, the Hamiltonian becomes
\begin{eqnarray}
H_C  &=& \int {d^3 x} \left\{ {c(x)  +
\frac{1}{2}{\bf B}^2 } \right\} \nonumber\\
&-& \int {d^3 x} \frac{1}{2}\Pi _i \left[ {1 - \frac{{g^2 {\bf v}^2 }}{4}
\frac{\Delta }{{\left( {\xi ^2 \Delta  + m^2 } \right)}}}
\right]\Pi ^i, \label{Tormag35}
\end{eqnarray}
where $c(x) = c_1 (x) - a_0 (x)$.

According to the usual procedure we introduce a supplementary condition
on the vector potential such that the full set of constraints becomes
second class. A particularly convenient choice is found to be
\begin{equation}
\Gamma _2 \left( x \right) \equiv \int\limits_{C_{\xi x} } {dz^\nu }
A_\nu \left( z \right) \equiv \int\limits_0^1 {d\lambda x^i } A_i
\left( {\lambda x} \right) = 0, \label{Tormag40}
\end{equation}
where  $\lambda$ $(0\leq \lambda\leq1)$ is the parameter describing
the spacelike straight path $ x^i = \xi ^i  + \lambda \left( {x -
\xi } \right)^i $, and $ \xi $ is a fixed point (reference point).
There is no essential loss of generality if we restrict our
considerations to $ \xi ^i=0 $. The choice (\ref{Tormag40}) leads
to the Poincar\'e gauge \cite{GaeteZ,GaeteSPRD}. As a consequence,
the only nontrivial Dirac bracket for the canonical variables is
given by
\begin{eqnarray}
\left\{ {A_i \left( x \right),\Pi ^j \left( y \right)} \right\}^ *
&=&\delta{ _i^j} \delta ^{\left( 3 \right)} \left( {x - y} \right)
\nonumber\\
&-& \partial _i^x \int\limits_0^1 {d\lambda x^j } \delta ^{\left( 3
\right)} \left( {\lambda x - y} \right). \label{Tormag45}
\end{eqnarray}

We are now equipped to compute the interaction energy for the model
under consideration. As mentioned before, in order to accomplish this
purpose we will calculate the expectation value of the energy operator
$H$ in the physical state $|\Phi\rangle$. Let us also mention here that,
as was first established by Dirac \cite{Dirac}, the physical
state $|\Phi\rangle$ can be written as
\begin{eqnarray}
\left| \Phi  \right\rangle &\equiv& \left| {\overline \Psi  \left(
\bf y \right)\Psi \left( {\bf y}\prime \right)} \right\rangle \nonumber\\
&=&
\overline \psi \left( \bf y \right)\exp \left(
{iq\int\limits_{{\bf y}\prime}^{\bf y} {dz^i } A_i \left( z \right)}
\right)\psi \left({\bf y}\prime \right)\left| 0 \right\rangle,
\label{Tormag50}
\end{eqnarray}
where the line integral is along a spacelike path on a fixed time
slice, and $\left| 0 \right\rangle$ is the physical vacuum state.
Notice that the charged matter field together with the electromagnetic
cloud (dressing) which surrounds it, is given by
$\Psi \left( {\bf y} \right) = \exp \left( { - iq\int_{C_{{\bf \xi}
{\bf y}} } {dz^\mu A_\mu  (z)} } \right)\psi ({\bf y})$. Thanks to
our path choice, this physical fermion then becomes $\Psi \left(
{\bf y} \right) = \exp \left( { - iq\int_{\bf 0}^{\bf y} {dz^i  }
A_{i} (z)} \right)\psi ({\bf y})$. In other terms, each of the
states ($\left| \Phi  \right\rangle$) represents a
fermion-antifermion pair surrounded by a cloud of gauge fields to
maintain gauge invariance.

Taking into account the above Hamiltonian structure, we observe that
\begin{eqnarray}
\Pi _i \left( x \right)\left| {\overline \Psi  \left( \bf y
\right)\Psi \left( {{\bf y}^ \prime  } \right)} \right\rangle  &=&
\overline \Psi  \left( \bf y \right)\Psi \left( {{\bf y}^ \prime }
\right)\Pi _i \left( x \right)\left| 0 \right\rangle \nonumber\\
&+&  q\int_ {\bf
y}^{{\bf y}^ \prime  } {dz_i \delta ^{\left( 3 \right)} \left( {\bf
z - \bf x} \right)} \left| \Phi \right\rangle. \nonumber\\
\label{Tormag55}
\end{eqnarray}
Having made this observation and since the fermions are taken to be
infinitely massive (static) we can substitute $\Delta$ by
$-\nabla^{2}$ in Eq. (\ref{Tormag35}). Therefore, the expectation
value $\left\langle H \right\rangle _\Phi$ is expressed as
\begin{equation}
\left\langle H \right\rangle _\Phi   = \left\langle H \right\rangle _0
+ \left\langle H \right\rangle _\Phi ^{\left( 1 \right)}  +
\left\langle H \right\rangle _\Phi ^{\left( 2 \right)}, \label{Tormag60}
\end{equation}
where $\left\langle H \right\rangle _0  = \left\langle 0
\right|H\left| 0 \right\rangle$. The $\left\langle H \right\rangle _\Phi ^{\left( 1 \right)}$ and $\left\langle H \right\rangle _\Phi ^{\left( 2 \right)}$
terms are given by
\begin{equation}
\left\langle H \right\rangle _\Phi ^{\left( 1 \right)}  =  - \frac{1}{2}
\frac{b}{{\xi ^2 }}\left\langle \Phi  \right|\int {d^3 x} \Pi _i
\frac{{\nabla ^2 }}{{\left( {\nabla ^2  - M^2 } \right)}}\Pi ^i \left|
\Phi  \right\rangle, \label{Tormag65a}
\end{equation}
and
\begin{equation}
\left\langle H \right\rangle _\Phi ^{\left( 2 \right)}  = \frac{{M^2 }}{2}
\left\langle \Phi  \right|\int {d^3 x} \Pi _i \frac{1}{{\left( {\nabla ^2  - M^2 } \right)}}\Pi ^i \left| \Phi  \right\rangle,  \label{Tormag65b}
\end{equation}
where $M^2  \equiv \frac{{m^2 }}{{\xi ^2 }} = \frac{{m^2 }}{{b +
g^2 {\cal B}^2 }}$. Using Eq. (\ref{Tormag55}), the $\left\langle H
\right\rangle _\Phi ^ {\left( 1 \right)}$ and $\left\langle H \right\rangle _\Phi ^
{\left( 2 \right)}$ terms can be rewritten as
\begin{eqnarray}
\left\langle H \right\rangle _\Phi ^{\left( 1 \right)}  &=&  - \frac{1}{2}
\frac{{bq^2 }}{{(b + g^2 {\cal B}^2)}}\int {d^3 x} \int_{\bf y}^{{\bf y}^
\prime  } {dz_i^ \prime  } \delta ^{\left( 3 \right)} \left( {{\bf x} -
{\bf z}^ \prime  } \right) \nonumber\\
&\times&\left( {1 - \frac{{\nabla ^2 }}{{M^2 }}}
\right)_x^{ - 1} \int_{\bf y}^{{\bf y}^ \prime  } {dz^i }
\delta ^{\left( 3 \right)} \left( {{\bf x} - {\bf z}} \right),
\label{Tormag70a}
\end{eqnarray}
and
\begin{eqnarray}
\left\langle H \right\rangle _\Phi ^{\left( 2 \right)} &=& \frac{{M^2q^2
}}{2}\int {d^3 } x\int_{\bf y}^{{\bf y}^{\prime}} {dz_i^{\prime }}
\delta ^{\left( 3 \right)} \left( {{\bf x} - {\bf z}^{\prime}  }
\right) \nonumber\\
&\times& \left( {\nabla ^2 - M^2 } \right)_x^{ - 1}
\int_{\bf y}^{{\bf y}^{\prime}} {dz^i } \delta ^{\left( 3 \right)}
\left( {{\bf x} - {\bf z}} \right).   \label{Tormag70b}
\end{eqnarray}
Following our earlier procedure \cite{GaeteS,GaeteS2}, we see that the
potential for two opposite charges located at ${\bf y}$ and ${\bf
y^{\prime}}$ takes the form
\begin{eqnarray}
V &=&  - \frac{{q^2 b}}{{4\pi \left( {b + g^2 {\cal B}^2 } \right)}}
\frac{{e^{ - \left( {\sqrt {\frac{{m^2 }}{{b + g^2 {\cal B}^2 }}} }
\right)L} }}{L} \nonumber\\
&+& \frac{{q^2 m^2 }}{{8\pi \left( {b + g^2 {\cal B}^2 }
\right)}}\ln \left( {1 + \frac{{\Lambda ^2 }}{{m^2 }}\left( {b +
g^2 {\cal B}^2 } \right)} \right)L,  \label{Tormag75}
\end{eqnarray}
where $\Lambda$ is a cutoff and $|{\bf y} -{\bf y}^{\prime}|\equiv
L$. Hence we see that the static potential profile displays a
confining behavior. Notice that expression (\ref{Tormag75}) is
spherically symmetric, although the external fields break the
isotropy of the problem in a manifest way. The Yukawa-type 
component to the potential above vanishes whenever $b=0$. It
actually signals the contribution of a scalar boson (the spin-$0$
mode of $S_\mu$) to the interparticle potential. And, from our
Lagrangian (\ref{Tormag5}), this scalar boson shows up if and 
only if $b\neq0$. So, this Yukawa screening appears as a byproduct
of a dynamical spin-$0$ torsion. As previously pointed out, for
$m^2 \ll m_0^2$, which is the physically acceptable regime, this
screening fades off.

It is now important to give a meaning to the cutoff $\Lambda$.
To do that, we should recall that our effective model for the
electromagnetic field is an effective description that comes out
upon integration over the torsion, whose excitation is massive.
$1/m$, the Compton wavelength of this excitation, naturally defines
a correlation distance. Physics at distances of the order or lower
than $1/m$ must necessarily take into account a microscopic description
of torsion. This means that, if we work with energies of the order or
higher than m, our effective description with the integrated effects
of $S^\mu$ is no longer sensible. So, it is legitimate that, for the
sake of our analysis, we identify $\Lambda$ with $m$. Then, with this
identification, the potential of Eq. (\ref{Tormag75})  takes the form
below:
\begin{eqnarray}
V &=&  - \frac{{q^2 b}}{{4\pi \left( {b + g^2 {\cal B}^2 } \right)}}
\frac{{e^{ - \left( {\sqrt {\frac{{m^2 }}{{b + g^2 {\cal B}^2 }}} }
\right)L} }}{L} \nonumber\\
&+& \frac{{q^2 m^2 }}{{8\pi \left( {b + g^2 {\cal B}^2 }
\right)}}\ln \left( {1 + b + g^2 {\cal B}^2 } \right)L. \label{Tormag80}
\end{eqnarray}
An immediate consequence of this is that for $b=0$ the screening term
(encoded in the Yukawa potential) disappears of the static potential
profile, describing an exactly confining phase \cite{GaeteHel09}.

\subsection{Electric case}

We now extend what we have done to the case $v^{0i}=0$ and
$v^{ij}\ne 0$ (referred to as the electric one in what follows). Thus,
the corresponding Lagrangian is given by
\begin{equation}
{\cal L}_{eff}  = - \frac{1}{4}f_{\mu \nu } f^{\mu \nu }  + \frac{{g^2 }}
{8}v^{ij} f_{ij} \frac{\Delta }{{\left( {b\Delta  + m^2 } \right)}}v^{kl} f_{kl}, \label{Tormag85}
\end{equation}
with $\mu ,\nu  = 0,1,2,3$ and $i,j,k,l = 1,2,3$. Following the same
steps employed for obtaining (\ref{Tormag75}), we now carry out a
Hamiltonian analysis of this model. First, the canonical momenta
following from Eq.(\ref{Tormag85}) are $\Pi^\mu=f^{\mu0}$,
which results in the usual primary constraint $\Pi^0=0$ and
$\Pi^i=f^{i0}$. Defining the electric and magnetic fields by $ E^i =
f^{i0}$ and $B^i  = -\frac{1}{2}\varepsilon ^{ijk} f_{jk}$,
respectively, the canonical Hamiltonian can be written as
\begin{eqnarray}
H_C  &=& \int {d^3 x} \left\{ { - A_0 \partial _i \Pi ^i  +
\frac{1}{2}{\bf \Pi} ^2  + \frac{1}{2}{\bf B}^2 } \right\} \nonumber\\
&-& \frac{{g^2 }}{{8}}\int {d^3 x} \left\{ {\varepsilon _{ijm}
\varepsilon _{k\ln } v^{ij} B^m \frac {\Delta}{(b \Delta + m^2)}
v^{kl} B^n } \right\}.\nonumber\\
\label{Tormag90}
\end{eqnarray}
Time conservation of the primary constraint leads to the secondary
constraint, $\Gamma_1(x) \equiv \partial_i\Pi^i=0$, and the time
stability of the secondary constraint does not induce more
constraints, which are first class. It should be noted that the
constrained structure for the gauge field is identical to the usual
Maxwell theory. Therefore, the corresponding expectation value is given
by
\begin{equation}
\left\langle H \right\rangle _\Phi   = \frac{1}{2}\left\langle \Phi
\right|\int {d^3 } x\Pi ^2 \left| \Phi  \right\rangle.
\label{Tormag95}
\end{equation}
As was explained in \cite{GaetePRD}, expression (\ref{Tormag95})
becomes
\begin{equation}
V =  - \frac{{q^2 }}{{4\pi }}\frac{1}{{\mid {\bf y} - {\bf y}^\prime
\mid }}, \label{Tormag100}
\end{equation}
which it is just the Coulomb potential.

Our understanding on the remarkable distinction between the cases of
an external electric and magnetic fields is as follows. Back to
the Lagrangian (\ref{Tormag5}), the torsion-electromagnetic field
interaction term can be also written as
\begin{equation}
{\cal L}_{int}  =  - g\left( {\partial _\lambda  S^\lambda  } \right)
{\bf E} \cdot {\bf B}, \label{Tormag105}
\end{equation}
up to a surface term. By splitting the classical background and the quantum
fluctuation of the electromagnetic field,
\begin{equation}
{\bf E} = \left\langle {\bf E} \right\rangle  + {\bf e}, \label{Tormag110a}
\end{equation}
and
\begin{equation}
{\bf B} = \left\langle {\bf B} \right\rangle  + {\bf b}, \label{Tormag110b}
\end{equation}
we readily see that:

(i) In the case the background is purely magnetic, $ \langle{\bf E}\rangle=0$,
torsion couples to the fluctuation ${\bf e}$ via $\langle {\bf B}\rangle$.
Since we are seeking the interparticle potential in the static regime  the
interaction term with $\langle{\bf B}\rangle$ present is 
$ - g\left( {\partial _\lambda  S^\lambda  } \right)\left\langle {\bf B} \right\rangle  \cdot {\bf e} =  - g\left( {\partial _\lambda  S^\lambda  } \right)\left\langle {\bf B} \right\rangle  \cdot {\bf \nabla} \varphi$, so that, we are lead to conclude that it is the interchange of the scalar $\varphi$ the responsible for the Yukawa-like piece of the potential.

(ii) For an electric background,  $ \langle{\bf B}\rangle=0$, the interaction term
with  $\langle {\bf E}\rangle$ present reads 
$- g\left( {\partial _\lambda  S^\lambda  } \right)\left\langle {\bf E} \right\rangle  \cdot {\bf b} =  - g\left( {\partial _\lambda  S^\lambda  } \right)\left\langle {\bf E} \right\rangle  \cdot \left( {{\bf \nabla}  \times {\bf A}} \right)$. In this case, the
contribution to the potential is due to an ${\bf A}$-exchange and this is why no
Yukawa-term shows up.

\section{Final Remarks}

In summary, by using the gauge-invariant but path-dependent
formalism, we have extended our previous analysis about the static
potential for a theory which includes both spin-$1$ and spin-$0$
states for the axial torsion field $S_\mu$ coupled to photons,
in the case when there are nontrivial constant expectation values
for the gauge field strength $F_{\mu\nu}$.

It was shown that in the case when $\langle F_{\mu\nu} \rangle$
is electric-like no unexpected features are found. Indeed, the
resulting static potential remains Coulombic. More interestingly,
it was shown that when $\langle F_{\mu\nu} \rangle$ is magnetic-like
the potential between static charges displays a Yukawa
piece plus a linear confining piece. We stress here the role played
by the spin-$0$ state of the torsion field $S_\mu$ in yielding the
Yukawa potential. An analogous static potential profile in axionic
electrodynamics may be recalled \cite{GaeteG}.   \\

\section{Acknowledgments}

One of us (PG) wants to thank the Field Theory Group of the CBPF for
hospitality. This work was supported in part by Fondecyt (Chile)
grant 1080260. I. L. Shapiro is kindly acknowledged
for discussions on dynamical torsion.

\end{document}